%% file: arxiv-mttkrp.tex
\documentclass[conference]{IEEEtran}
\IEEEoverridecommandlockouts

\usepackage{amsmath,amssymb,amsfonts}
\usepackage{graphicx}
\usepackage{textcomp}
\usepackage{xcolor}
\usepackage{algorithm}
\usepackage[hyphens]{url}
\usepackage[switch]{lineno}
\usepackage[colorlinks,
            bookmarksnumbered,
            citecolor=black,
            urlcolor=black,
            linkcolor=black
            ]{hyperref}
\usepackage[capitalize]{cleveref}
\def\BibTeX{{\rm B\kern-.05em{\sc i\kern-.025em b}\kern-.08em
    T\kern-.1667em\lower.7ex\hbox{E}\kern-.125emX}}
\usepackage[numbers]{natbib}
\usepackage[inline]{enumitem}
\usepackage{pgf}
\input{macros/table}

\input{macros/setup}

\input{macros/alg}
\input{macros/math}

\begin{document}

\title{A Performance Portable Matrix Free Dense MTTKRP in GenTen
\thanks{
Sandia National Laboratories is a multimission laboratory managed and operated by National Technology \& Engineering Solutions of Sandia, LLC, a wholly owned subsidiary of Honeywell International Inc., for the U.S. Department of Energy’s National Nuclear Security Administration under contract DE-NA0003525.
This paper describes objective technical results and analysis. Any subjective views or opinions that might be expressed in the paper do not necessarily represent the views of the U.S. Department of Energy or the United States Government.
SAND2025-13211O.
}
}

\author{\IEEEauthorblockN{Gabriel Kosmacher}
\IEEEauthorblockA{\textit{Oden Institute} \\
\textit{The University of Texas at Austin}\\
Austin, TX USA \\
gkosmacher@utexas.edu}
\and
\IEEEauthorblockN{Eric T. Phipps}
\IEEEauthorblockA{\textit{Sandia National Laboratories} \\
Albuquerque, NM USA\\
etphipp@sandia.gov}
\and
\IEEEauthorblockN{Sivasankaran Rajamanickam}
\IEEEauthorblockA{\textit{Sandia National Laboratories} \\
Albuquerque, NM USA\\
srajama@sandia.gov}
}

\maketitle

\input{sections/abstract}

\begin{IEEEkeywords}
tensor decomposition, canonical polyadic, MTTKRP, Kokkos, GPU
\end{IEEEkeywords}

\input{sections/introduction}

\input{sections/background}

\input{sections/methods}

\input{sections/results}
\input{sections/conclusion}

\section*{Acknowledgment}
 This material is based upon work partially supported by the U.S. Department of Energy, Office of Science, Office of Advanced Scientific Computing Research under contract number DE-NA0003525.

\bibliographystyle{IEEEtranN}
\bibliography{ref}

\end{document}

%% file: macros/table.tex
\usepackage{
	colortbl, 
	booktabs, 
	multirow,
	siunitx,
	makecell,
}
\newcolumntype{g}{>{\columncolor{gray!20}}c}

%% file: macros/setup.tex
\usepackage[mathscr]{eucal}

\usepackage{stmaryrd}

\usepackage{amsbsy}

\usepackage{bm}

\usepackage{braket}

\usepackage{xspace}

\usepackage{xparse}

\usepackage{algpseudocode}

\usepackage{lmodern}
\usepackage{listings}
\usepackage[most]{tcolorbox}

\DeclareTCBListing{macrobox}{s G{} }{IfBooleanTF={#1}{}{listing side text},title=#2,
  listing options={style=tcblatex,commentstyle=\color{red!70!black}}
}

\usepackage{fixme}
\fxsetup{
  status=draft,
  nomargin,
  inline,
  theme=color,
}
\fxsetup{marginface=\linespread{1}\footnotesize}
\FXRegisterAuthor{tk}{tke}{TK}
\FXRegisterAuthor{nj}{nje}{NJ}
\FXRegisterAuthor{ep}{epe}{EP}

\NewDocumentCommand{\VcRoot}{m m G{\BooleanFalse} G{\BooleanFalse}}{%
    \bm{#1\mathbf{\MakeLowercase{#2}}}%
    \IfBooleanT{#3}{^{\intercal}}%
    \IfBooleanT{#4}{^{\phantom{\intercal}}}%
  }
\NewDocumentCommand{\Vc}{ O{} m t' t"} {\VcRoot{#1}{#2}{#3}{#4}}

\NewDocumentCommand{\MxRoot}{m m G{\BooleanFalse} G{\BooleanFalse}}{%
    \bm{#1\mathbf{\MakeUppercase{#2}}}%
    \IfBooleanT{#3}{^{\intercal}}%
    \IfBooleanT{#4}{^{\phantom{\intercal}}}%
  }
\NewDocumentCommand{\Mx}{O{} m t' t"}{\MxRoot{#1}{#2}{#3}{#4}}

\NewDocumentCommand{\Tn}{O{} m}{\boldsymbol{#1\mathscr{\MakeUppercase{#2}}}}

\NewDocumentCommand{\Tm}{O{} m m t' t"}{\MxRoot{#1}{#2}{#4}{#5}_{(#3)}}

\NewDocumentCommand{\X} {} {\Tn{X}}

\NewDocumentCommand{\Xk} { O{} G{(k)} t' t"} {\MxRoot{#1}{X}{#3}{#4}_{#2}}

\DeclareDocumentCommand{\xi}{ s } 
{
  \IfBooleanTF{#1}
  {x_{i_1 \dots i_d}}
  {x_{i}}
}

\def\lastiter{\bar}

\NewDocumentCommand{\Xt} {s} {\IfBooleanTF{#1}{\Mx{X}_t}{\Tn{X}_t}}
\NewDocumentCommand{\Xh} {s} {\IfBooleanTF{#1}{\Mx{X}_h}{\Tn{X}_h}}

\NewDocumentCommand{\yit}{s}{\IfBooleanT{#1}{\tilde}y_{it}}

\NewDocumentCommand{\yiht}{s}{\IfBooleanT{#1}{\tilde}y_{ih}}
\NewDocumentCommand{\cit}{s}{\IfBooleanT{#1}{\tilde}c_{it}}

\NewDocumentCommand{\Wt} {s} {\IfBooleanTF{#1}{\Mx{W}_t}{\Tn{W}_t}}
\NewDocumentCommand{\Wh} {s} {\IfBooleanTF{#1}{\Mx{W}_h}{\Tn{W}_h}}

\NewDocumentCommand{\M} {} {\Tn{M}}
\NewDocumentCommand{\Mt} {s} {\IfBooleanTF{#1}{\Mx{M}_t}{\Tn{M}_t}}
\NewDocumentCommand{\Mtold} {s} {\IfBooleanTF{#1}{\Mx[\lastiter]{M}_t}{\Tn[\lastiter]{M}_t}}
\NewDocumentCommand{\Mh} {s} {\IfBooleanTF{#1}{\Mx{M}_h}{\Tn{M}_h}}

\NewDocumentCommand{\Mk} { O{k} } {\mathbf{M}_{(#1)}}

\NewDocumentCommand{\A}{t'}{\mathbf{A}\IfBooleanT{#1}{^{\mkern -4mu \intercal}}}
\NewDocumentCommand{\B}{}{\Mx{B}}
\NewDocumentCommand{\C}{}{\Mx{C}}

\NewDocumentCommand{\Fac} { G{A}G{k} } {%
  \Mx{#1}_{#2}
}
\NewDocumentCommand{\Ak} { G{k} } {%
  \Fac{A}{#1}
}

\NewDocumentCommand{\akj}{O{k} G{j}}{%
  \Vc{a}^{\mkern -3mu {(#1)}}_{#2}
}

\NewDocumentCommand{\Akold} { G{k} } {%
  \Mx[\lastiter]{A}^{\mkern -4mu (#1)}
}

\NewDocumentCommand{\lj}{G{j} t"}{\lambda_{#1}\IfBooleanT{#2}{^{\phantom{(1)}}\!\!}}
\NewDocumentCommand{\lvec}{}{\Vc{\lambda}}

\NewDocumentCommand{\st}{G{t} t'}{\Vc{s}_{#1} \IfBooleanT{#2}{^{\intercal}}}

\NewDocumentCommand{\HWt}{ G{t} }{\mathcal{H}_{#1}}

\NewDocumentCommand{\Y}{s}{\Tn[\IfBooleanT{#1}{\tilde}]{Y}}
\NewDocumentCommand{\Yt}{s}{\Tn[\IfBooleanT{#1}{\tilde}]{Y}_t}
\NewDocumentCommand{\Yh}{s}{\Tn[\IfBooleanT{#1}{\tilde}]{Y}_h}
\NewDocumentCommand{\Yht}{s}{\Tn[\IfBooleanT{#1}{\tilde}]{Y}_{h}}
\NewDocumentCommand{\Yk}{s}{\Mx[\IfBooleanT{#1}{\tilde}]{Y}_{\mkern -3mu (k)}}
\NewDocumentCommand{\Yn} { O{} G{n} t' t"} {\MxRoot{#1}{Y}{#3}{#4}_{#2}}
\NewDocumentCommand{\Ytk}{s}{(\Tn[\IfBooleanT{#1}{\tilde}]{Y}_t)_{\mkern -3mu (k)}}
\NewDocumentCommand{\Yhk}{s}{(\Yh)_{(k)}}
\NewDocumentCommand{\Yhtk}{s}{(\Yht)_{(k)}}
\NewDocumentCommand{\Yv}{s}{\Vc[\IfBooleanT{#1}{\tilde}]{Y}}

\NewDocumentCommand{\KT} { s } {
  \llbracket 
  \IfBooleanTF{#1}{\lvec;}{}
  \Ak{1}, \dots,  \Ak{d} \rrbracket
}

\NewDocumentCommand{\KTt} { G{\st} } {
  \llbracket 
  #1;
  \Ak{1}, \dots,  \Ak{d} \rrbracket
}
\NewDocumentCommand{\KTtt} { G{\st}G{t} } {
  \llbracket 
  #1;
  \Akt{1}{#2}, \dots,  \Akt{d}{#2} \rrbracket
}

\NewDocumentCommand{\KTdef}{O{} O{} m !g}{%
  #1\llbracket \IfValueTF{#4}{#3;#4}{#3} #2\rrbracket%
}
\NewDocumentCommand{\KTauto}{m !g}{\KT[\left][\right]{#1}{#2}}
\NewDocumentCommand{\KTbig}{m !g}{\KT[\bigl][\bigr]{#1}{#2}}
\NewDocumentCommand{\KTbigg}{m !g}{\KT[\biggl][\biggr]{#1}{#2}}
\NewDocumentCommand{\KTBig}{m !g}{\KT[\Bigl][\Bigr]{#1}{#2}}
\NewDocumentCommand{\KTBigg}{m !g}{\KT[\Biggl][\Biggr]{#1}{#2}}

\NewDocumentCommand{\Gk} { G{k} } {%
  \Mx{G}^{(#1)}
}

\NewDocumentCommand{\Zk} { G{k} t'} {\MxRoot{}{Z}{#2}_{#1}}

\NewDocumentCommand{\Zkold} { G{k} t'} {\MxRoot{}{Z}{#2}_{#1}}

\NewDocumentCommand{\Z} { t' } {\MxRoot{}{Z}{#1}}

\NewDocumentCommand{\D} {} {\mathcal{D}}

\NewDocumentCommand{\FPD} { s m m } {
  \IfBooleanTF{#1}
  {\partial #2 / \partial #3}
  {\frac{\partial #2}{\partial #3}}
}

\DeclareMathOperator{\diag}{diag}

%% file: macros/alg.tex
\algrenewcommand\algorithmicrequire{\textbf{Input:}}
\algrenewcommand\algorithmicensure{\textbf{Output:}}

\algblockdefx[LeagueRange]{LeagueRange}{EndLeagueRange}[1][]{\textbf{LeagueRange} #1}{\textbf{End LeagueRange}}
\algblockdefx[TeamRange]{TeamRange}{EndTeamRange}[1][]{\textbf{TeamRange} #1}{\textbf{End TeamRange}}
\algblockdefx[VecRange]{VecRange}{EndVecRange}[1][]{\textbf{VecRange} #1}{\textbf{End VecRange}}
\algblockdefx[SIMD]{SIMD}{EndSIMD}[1][]{\textbf{SIMDRange} #1}{\textbf{End SIMDRange}}

\algrenewcommand{\algorithmiccomment}[1]{\hfill \textcolor{gray}{// #1}}

%% file: macros/math.tex
\newcommand{\R}{\mathbb{R}}
\newcommand{\bldl}{\boldsymbol{\lambda}}

\newcommand{\flops}{\mathbf{f}}
\newcommand{\mops}{\mathbf{m}}
\newcommand{\ai}{\mathbf{I}}
\newcommand{\thru}{\mathbf{gflops}}
\newcommand{\band}{\mathbf{mops}}

%% file: sections/abstract.tex
\begin{abstract}
We extend the GenTen tensor decomposition package by introducing an accelerated dense matricized tensor times Khatri-Rao product (MTTKRP), the workhorse kernel for canonical polyadic (CP) tensor decompositions, that is portable and performant on modern CPU and GPU architectures.
In contrast to the state-of-the-art matrix multiply based MTTKRP kernels used by Tensor Toolbox, TensorLy, etc., that explicitly form Khatri-Rao matrices,
we develop a matrix-free element-wise parallelization approach whose memory cost grows with the rank $R$ like the sum of the tensor shape $\mathcal{O}(R(n+m+k))$, compared to matrix-based methods whose memory cost grows like the product of the tensor shape $\mathcal{O}(R(mnk))$.
For the largest problem we study, a rank 2000 MTTKRP, the smaller growth rate yields a matrix-free memory cost of just $2\%$ of the matrix-based methods, a $50\times$ improvement.
In practice, the reduced memory impact means our matrix-free MTTKRP can compute a rank 2000 tensor decomposition on a single NVIDIA H100 instead of six H100s using a matrix-based  MTTKRP.
We also compare our optimized matrix-free MTTKRP to baseline matrix-free implementations on different devices, showing a $2\times$ single-device speedup on an Intel 8480+ CPU, an $11\times$ speedup on an NVIDIA H100 GPU and a $6\times$ speedup on an AMD MI300A GPU.
In addition to numerical results, we provide fine grained performance models for an ideal multi-level cache machine, compare analytical performance predictions to empirical results, and provide a motivated heuristic selection for selecting an algorithmic hyperparameter.
\end{abstract}

%% file: sections/introduction.tex
\section{Introduction}
Dense tensors are relevant in the fields of numerical simulations \cite{goal_decomp}, medical imaging \cite{human_connection}, signal processing \cite{sig_proc}, and color perception \cite{color}, among others.
Low-rank approximations of dense tensors are powerful tools used to compress such datasets and reveal relationships between modes.
A popular decomposition for low-rank approximations is the CANDECOMP/PARAFAC (CP) decomposition, also called canonical polyadic decomposition \cite{tens_sirev}; the decomposition takes as input a user-supplied rank $R$ and outputs a weighted sum of $R$ rank-1 tensors.
The bottleneck kernel for CP algorithms \cite{ttb} is the matrix action of a mode-$k$ unfolding of a tensor with the Khatri-Rao product of factor matrices and is called the \emph{matricized tensor times Khatri-Rao product} (MTTKRP).

Most open-source tensor decomposition packages (see \cref{sec:related}) compute the MTTKRP by explicitly forming the Khatri-Rao product matrix, allowing invocations of BLAS-3 algorithms, but \emph{compromising memory usage}.
In contrast, we seek to compute the MTTKRP \emph{without} explicitly forming the Khatri-Rao matrix and \emph{instead} take advantage of the problem structure to develop \emph{matrix-free} (also called \emph{element-based}) algorithms.
Our work extends the open-source GenTen\footnote{\url{https://github.com/sandialabs/GenTen}} tensor decomposition package with an implementation of our dense matrix-free MTTKRP and, \emph{as far as we know, introduces the first GPU algorithm for a matrix-free MTTKRP with dense tensors}.

\textbf{Contributions:} Our work offers the following methodological contributions and experimental results:
\begin{itemize}
	\item The design and open-source implementation of performance-portable matrix-free algorithms for dense MTTKRPs (see \cref{sec:methods}). For a tensor $\Y$ of shape $I_1, \dots, I_d$ and CP-rank $R$, our matrix-free algorithms for a mode-$k$ MTTKRP have $\mathcal{O}(R\sum_nI_n)$ memory usage, while standard matrix-based approaches have $\mathcal{O}(R\prod_{n\neq k}I_n)$ memory cost.
	\item Multiple analytical memory and computational models for our algorithmic variants based on an ideal multi-level cache machine with different caching assumptions (see \cref{sec:perf}).
	\item A heuristic model for an algorithmic hyper-parameter that aligns with measured times on a CPU and GPU (see \cref{sec:tilesize}).
	\item CPU and GPU algorithmic evaluations (see \cref{sec:single,sec:dist}). In particular, our matrix-free MTTKRP can perform a tensor decomposition on one GPU that requires six GPUs for a matrix-based MTTKRP. We also compare our optimized, matrix-free MTTKRP to baseline matrix-free implementations, showing a $2\times$ speedup on an Intel 8480$+$ CPU, an 11$\times$ speedup on an NVIDIA H100 GPU, and a $6\times$ speedup on an AMD MI300A GPU.
\end{itemize}

\subsection{Related work}\label{sec:related}
A variety of open-source packages offer dense CP decompositions. N-Way Toolbox \cite{n_way}, TensorLy \cite{tensorly}, and rTensor \cite{rtensor} compute MTTKRPs by explicitly unfolding the tensor and forming the Khatri-Rao product.
The MATLAB Tensor Toolbox \cite{ttb, dense_ttb}, PyTTB \cite{pyttb} and GenTen \cite{genten_code} use an algorithm introduced by \citet{phan_gemm}, described in \cref{sec:methods-gemm}, that avoids expensive tensor unfolding and reduces memory impact by forming \emph{partial} Khatri-Rao matrices.
\citet{block_gemm} build upon \cite{phan_gemm} by constructing an algorithm, implemented using the Eigen3 C++ library \cite{eigenweb}, that sequentially stores blocks of the Khatri-Rao matrix in constant memory, hence achieving the desired $\mathcal{O}(R\sum_nI_n)$ storage complexity.
\emph{However}, the blocking algorithm is only applicable to the special case where all $k$-mode MTTKRPs are computed at once, and cannot generalize to compute individual mode-$k$ MTTKRPs needed by, e.g., the CP-ALS algorithm. 
\cref{tab:codes} gives an overview of features for different dense mode-$k$ MTTKRP algorithms and codebases.

\begin{table*}[h]
\centering
\caption{Methods and codebases for dense mode-$k$ MTTKRPs. Given a tensor $\Y$ of shape $I_1,\dots,I_d$ with factor matrices $\A_1, \dots, \A_d$, let the Khatri-Rao matrix be $\Z = \Ak{d} \odot \dots \odot \Ak{k+1} \odot \Ak{k-1} \odot \cdots \odot \Ak{1}$ (see \cref{sec:def_mttkrp} for details). We list memory usage and open-source codes for each method, and we list CPU and GPU support for each codebase. All methods have the same asymptotic computational cost of $\mathcal{O}(Rd(\prod_n I_n))$.}
\begin{tabular}{ lccccc  }
 \toprule
 methods        & memory & code & CPU & GPU \\
 \midrule
 \multirow{3}*{full $\Z$ \cite{ttb}} &\multirow{3}*{$\mathcal{O}\left(R\prod\limits_{n \neq k}I_n\right)$} 
				       &N-way Toolbox \cite{n_way} & \checkmark \\
				       && TensorLy \cite{tensorly} & \checkmark & \checkmark\\
				       && rTensor \cite{rtensor}   & \checkmark\\
 \midrule
 \multirow{3}*{partial $\Z$ \cite{phan_gemm}} & \multirow{3}*{$\mathcal{O} \left(R\left(\prod\limits_{n=1}^{k-1} I_n + \prod\limits_{k+1}^d I_n\right) \right)$} & TTB \cite{ttb} & \checkmark \\ 
						&&PyTTB \cite{pyttb} & \checkmark \\ 
						&&GenTen \cite{genten_code} & \checkmark & \checkmark \\ 
 \midrule
 no $\Z$ (our method)  & $\mathcal{O}\left(R\sum\limits_{n=1}^dI_n\right)$ & GenTen & \checkmark & \checkmark\\ 
 \bottomrule
\end{tabular}
\label{tab:codes}
\end{table*}

There has been considerably more work on optimizing MTTKRPs in the sparse tensor case. 
DeFacTo \cite{defacto} stores tensors as collections of sparse matrices and relies on optimized sparse matrix-vector multiply routines to compute the MTTKRP.
HiCOO \cite{hicoo} proposes a new storage format that partitions the sparse tensor into sparse blocks to increase memory bandwidth and develops an algorithm that parallelizes over groups of sparse blocks.  
Both SPLATT and HiCOO were developed primarily for low-thread-count CPUs and do not consider portability to high-thread GPU architectures. 
\citet{alto} introduce another sparse tensor format, ALTO, that linearizes the tensor's multi-index set such that tensor entries close in space are close in memory. 
Entries are then decomposed into \emph{line segments} that encode subspaces, and a parallel algorithm is introduced that assigns threads to chunks of line segments.
Though developed for CPUs, increasing the number of chunks per line segment \emph{should} allow for straightforward portability to GPUs. 
GenTen \cite{genten} offers a performance-portable approach that  parallelizes over tensor nonzeros and permutes the nonzeros to avoid atomic contention.  
This algorithm is similar to ours (in the dense case) as it avoids forming the Khatri-Rao product explicitly. 
However, permuting the tensor nonzeros requires an additional storage cost of $\mathcal{O}(dN)$, where $d$ is the dimension of the tensor and $N$ is the number of nonzeros. As we show in \cref{sec:perf}, explicitly permuting tensor entries in such a fashion is \emph{not} necessary for an efficient dense MTTKRP.

%% file: sections/background.tex
\section{Background}

\subsection{Notation}\label{sec:not}
A tensor $\X \in \R^{I_1 \times \cdots \times I_d}$ is a $d$-way array of size $N=I_1 \times \dots \times I_d$ typically stored in a flattened row-major or column-major format.
It is standard practice \cite{tens_sirev} to denote tensors as bold uppercase letters in Euler calligraphic font (e.g., $\X$), matrices by bold uppercase letters (e.g., $\A$), vectors by bold lowercase letters (e.g., $\boldsymbol{a}$), and scalars by lowercase letters (e.g., $a$). 
We adopt MATLAB notation for array indexing\footnote{\url{https://www.mathworks.com/help/matlab/math/array-indexing.html}}.

The Khatri-Rao product $\odot$ of two matrices $\A \in \R^{m \times p}$ and $\B \in \R^{n \times p}$ is the column-wise Kronecker product of $\A$ and $\B$ defined by
\begin{equation}
	\A \odot \B = [\Vc{a}_1 \otimes \Vc{b}_1 | \dots | \Vc{a}_p \otimes \Vc{b}_p] \in \R^{mn \times p},
\end{equation}
where $\otimes$ is the Kronecker product and $\akj \equiv \Ak(:, j)$ is a column vector.

Tensors are indexed via multi-indexed tuples $i = (i_1, \dots, i_d)$ where $i_n \in [I_n]$. The set notation $[a]$ is shorthand for $\{z : z \in \mathbb{Z}, 1 \leq z \leq a\}$.
We call $i \in [N]$ the \emph{linear index} of a tensor and $(i_1, \dots, i_d)$ the \emph{multi-index}, we assume a bijective mapping between the two index sets, and we refer to a \emph{tensor element} as $x_i \equiv \X(i) \equiv \X(i_1, \dots, i_d)$. The forward map $i \mapsto (i_1, \dots, i_d)$ is given by the operator $\Call{ind2sub}{\X, i}$ and is defined like the MATLAB function of the same name\footnote{\url{https://www.mathworks.com/help/matlab/ref/ind2sub.html}}, while the backward map $(i_1, \dots i_d) \mapsto i$ is given by the operator $\Call{sub2ind}{\X, (i_1, \dots, i_d)}$, again defined like the corresponding MATLAB function\footnote{\url{https://www.mathworks.com/help/matlab/ref/sub2ind.html}}.
The $n$th mode-$k$ \emph{subtensor} of a tensor $\Tn{S}^{(k)}_n$ is a $d-1$ tensor obtained by fixing the $k$th value of the multi-index to $n$ and allowing the other values to range, i.e., $\Tn{S}^{(1)}_n = \X(n, :, \dots, :)$.
The mode-$k$ \emph{matricization} of a tensor $\X$ rearranges the tensor elements into a matrix $\Xk \in \R^{I_k \times N/I_k}$ such that element $i$ maps to $(i_k, i_k')$ by
\begin{equation}
	i_k' = 1 + \sum_{\substack{\ell=1 \\ \ell \neq k}}^{d} (i_\ell-1) \prod_{\substack{m=1 \\ m \neq k}}^{\ell-1}I_m.
\end{equation}
The \emph{reshape} operator $\Call{reshape}{\X, [s_1, \dots, s_m]}$ \emph{does not permute} the underlying tensor elements (unlike tensor matricization) and is defined like the MATLAB function of the same name\footnote{\url{https://www.mathworks.com/help/matlab/ref/double.reshape.html}}. 

The rank-$R$ \emph{canonical polyadic (CP) decomposition} \cite{Carr1970, parafac} of a tensor $\X$ is a rank-$R$ tensor $\M$ of the form
\begin{equation}
	\X \approx \M = \sum_{j=1}^R \lambda_j \Vc{a}_j^{(1)} \circ \dots \circ \Vc{a}_j^{(d)},
\end{equation}
where $\Vc{\lambda} \in \R^R$ is a weight vector and $\circ$ is a $d$-way outer product.
We call $\Ak = [\Vc{a}_1^{(1)}, \dots, \Vc{a}_d^{(k)}] \in \R^{I_k \times R}$ the mode-$k$ \emph{factor matrix} of a tensor $\X$. The form $\M = \KT$ is referred to as a \emph{Kruskal tensor}.

\subsection{The MTTKRP kernel}\label{sec:def_mttkrp}
Algorithms to compute CP decompositions generally fall into two categories: all-at-once \cite{cp-opt} or alternating \cite{cp_als_1, parafac}.
In both cases, the workhorse kernel is the mode-$k$ \emph{matricized tensor times Khatri-Rao product} (MTTKRP), defined as
\begin{equation}\label{eq:krp}
	\Gk = \Yk\Z\diag(\lvec),
\end{equation}
where $\Z = \Ak{d} \odot \dots \odot \Ak{k+1} \odot \Ak{k-1} \odot \cdots \odot \Ak{1}$ for some Kruskal tensor $\M=\KT$, $d$-way tensor $\Y$, and weight vector $\lvec \in \R_+^{R}$.
We call algorithms that explicitly construct $\Z$ \emph{matrix-based} MTTKRPs. 

Matrix-based MTTKRPs have two distinct downsides: 
\begin{enumerate*}[label=(\arabic*)]
    \item the explicit formation of the Khatri-Rao matrix $\Z$, and
    \item the explicit matricization $\Yk$. 
\end{enumerate*}
For a mode-$k$ MTTKRP with CP-rank $R$, the Khatri-Rao matrix $\Z$ is size $\prod_{n \neq k} I_n \times R$, which can dwarf the cost $R\sum_{n\neq k} I_n$ of simply storing the factor matrices $\A_n$ individually, leading to $\mathcal{O}(R\prod_{n \neq k} I_n)$ memory usage.  
As for the matricization $\Yk$, tensors are typically stored in memory as a flat array, and in such cases, the mode-$k$ (for $1 < k < d$) matricization requires reordering the non-contiguous tensor entries into contiguous chunks. 
These reorderings have a memory-bound cost of $\mathcal{O}((d-1)N)$ and impose significant costs on an otherwise computationally bound kernel.

These drawbacks motivate a \emph{matrix-free}, or \emph{element-wise}, definition of the mode-$k$ MTTKRP (common in the \emph{sparse} case \cite{genten}), requiring only the explicit storage of $\Y$, $\lvec$, and $\{\A_m \}_{m=1}^d$, given by
\begin{equation}\label{eq:elem}
	\Gk(n,j) = \lj \sum_{\substack{i=1 \\ i_k = n}}^N \Y(i) \prod_{\substack{m=1 \\ m \neq k}}^d \A_m(i_m,j),
\end{equation}
for $n \in [I_k]$ and $j \in [R]$. 
As the CP-rank $R$ grows, the dominant memory cost of this approach becomes the storage of the factor and output matrices, yielding an overall $\mathcal{O}(R\sum_n I_n)$ memory burden.  

\subsection{Kokkos programming framework}
Kokkos \cite{kokkos} is a C++ framework providing abstractions to write \emph{portable} functions for
single-instruction multiple-data (SIMD) parallelism on CPUs and single-instruction multiple-threads (SIMT) parallelism on GPUs.
It is the Kokkos convention to discuss levels of parallelism mirroring those of the OpenMP 4.0 specification \cite{omp}, i.e., \emph{leagues} of \emph{teams} comprising team-threads which may execute \emph{vector-thread} instructions in parallel.
Leagues are virtual and correspond to the highest level of parallel space (e.g., a CUDA grid or a collection of OpenMP threads),
while a team within a league corresponds to hyperthreads on a CPU and $y$-dimension threads within a block on a GPU.  Vector parallelism corresponds to vector instructions on CPUs and $x$-dimension threads (i.e., threads within a warp) on GPUs.
We use $\bf{LeagueRange}$, $\bf{TeamRange}$, and $\bf{VectorRange}$ to denote league, team, and vector parallel ranges, respectively.
We use $b_x$ to denote team size and $b_y$ to denote vector size, and we use $p_x$ to denote threads within a team
and $p_y$ to denote vector-threads within a thread.

\subsubsection{SIMD arrays}
Compile-time polymorphic SIMD arrays were introduced in \cite{genten} as an extension of Kokkos that allow for the allocation of small arrays as register arrays on GPUs or thread-private stack arrays on CPUs.
We utilize these arrays to allow for a single vector-thread $p_y$ to calculate many (e.g., $4$) updates to the output matrix $\Gk$ per vector loop, increasing memory-bandwidth efficiency. We denote the SIMD vector size as $F$ and SIMD vectors as Greek letters with the vector arrow, e.g., $\vec{\pi}, \vec{\varphi}$. 
SIMD vectors are in effect for loops that the compiler is forced to unroll and vectorize in registers.

%% file: sections/methods.tex
\section{Parallel MTTKRP algorithms}\label{sec:methods}
The state-of-the-art dense MTTKRP was introduced in \cite{phan_gemm} and takes the viewpoint of \cref{eq:krp}.
The algorithm, called $\Call{MTTKRP-GEMM}{}$ (\cref{alg:gemm}), uses strategic partitioning of the Khatri-Rao product matrix $\Z^T$ to avoid costly reorders of $\Y$, utilizes temporary memory allocations for a (sometimes) lower storage cost, and gets its name from its use of the BLAS-3 general matrix-matrix multiplies (GEMMs) for highly optimized parallelization.

We introduce element based algorithms all taking the viewpoint of \cref{eq:elem};
each of our algorithms utilize the \emph{same} vector parallelization over columns of $\{ \Ak \}$ and \emph{differ} in how they assign tensor elements $\Y(i)$ (or groups of $\Y(i)$'s) to Kokkos teams and threads, and are described as follows:
\begin{itemize}
       \item {\bf Element-ordered} (\cref{alg:elem}): assign each tensor element $\Y(i)$ to a thread; convert linear index to multi-index to read from $\{\Ak\}$; use atomics to resolve \emph{element-level} conflicts in $\Gk$.
       \item {\bf Subtensor-ordered} (\cref{alg:tile} with $N_T = N/I_k$): assign each subtensor $\Tn{S}^{(k)}_n$ to a team; read from $\{\Ak\}$ with column reuse; compute $\Gk$ in registers and write to global memory conflict free.
       \item {\bf Tile-ordered} (\cref{alg:tile}): assign subsets of each subtensor $\Tn{S}^{(k)}_n((t-1)N_T:tN_T)$, called tiles, to a team; read from $\{\Ak\}$ as in the subtensor-ordering algorithm; write to $\Gk$ with atomics to resolve \emph{tile-level} conflicts.
\end{itemize}
Each algorithm (including GEMM) takes as an input the tensor $\Y$, the set of factor matrices $\{\Ak\}_{m=1}^d$, the number of ranks $R$, the target mode $k$ (for a mode-$k$ MTTKRP); the tile-ordered algorithm takes in a hyperparameter $N_T$, called the tile volume; each matrix-free algorithms also take in a SIMD size $F$.

\Call{MTTKRP-ELEM}{}, the element-ordered algorithm, is parallel over the tensor elements $\Y(i)$ and is easy to implement, but has no structure when reading $\{\Ak\}$, yielding poor cache management and incurring $\mathcal{O}(NR)$ atomic writes to $\Gk$.

\Call{MTTKRP-SUB}{}, the subtensor-ordered algorithm, assigns each subtensor to a team---taking advantage of the fact that the $n$th row of $\Gk$ is determined solely by the tensor elements in $\Tn{S}^{(k)}_n$---to cache and reuse columns of $\{\Ak\}$ and write to $\Gk$ conflict free. However, assigning each subtensor to a team reduces parallelism when the subtensor size is larger than the team size as each thread must process many elements. 

\Call{MTTKRP-TILE}{}, the tile-ordered algorithm, attempts to resolve this issue by assigning teams to constant-sized subsets of subtensors, called tiles, allowing for more parallel threads to launch and ensuring a uniform workload across teams. Columns of $\{\Ak\}$ are reused as in the subtensor ordering, but $\Gk$ must be written with $\mathcal{O}{((N_S/N_T)R)}$ atomics---where $N_T$ is the number of elements in a tile and $N_S=N/I_k$ the number of elements in each subtensor---as different tiles in the same subtensor write to the same row; this ordering increases the parallelism compared to \Call{MTTKRP-SUB}{} and cache reuse and it decreases atomics compared to \Call{MTTKRP-ELEM}{}.

The vector parallelism scheme is the same as \cite{genten}: parallelism is introduced over the columns of the factor matrices $\{\Ak\}$ by assigning multiple column indices $j$ in \cref{eq:elem} to a thread vector $p_y$. 
Along with the enforcement of a row-wise layout, this allows for coalesced reads of the factor matrices, allowing the compute device to achieve a higher percentage of memory bandwidth. 
For simplicity, we assume that the CP-rank $R$ divides the product of the vector size and the SIMD array length $(b_y \times F)$ for the remainder of this discussion.

\subsection{MTTKRP-GEMM}\label{sec:methods-gemm}
The key insight of \cite{phan_gemm} is to avoid the costly matricizations $\Yk$ by using the \Call{reshape}{} operator and breaking up the \emph{full} Khatri-Rao product $\Z^T$ into \emph{left} and \emph{right} components $\Z^T = [\Z_R^T \in \R^{R \times I_R}| \Z_L^T \in \R^{\R \times I_L}]$, where $I_L = \prod_{m=1}^{k-1} I_m$ and $I_R = \prod_{m=k+1}^d I_m$. 

We briefly summarize the algorithm given in \cite{phan_gemm}. For more detail, please refer to the paper. For mode-$1$, replace $\Yk$ with $\Call{reshape}{\Y, [I_1, I_R]}$ in \cref{eq:krp} to avoid permutations. Similarly, for mode $d$, replace $\Yk$ with $\Call{reshape}{\Y, [I_L, I_d]}^T$ in \cref{eq:krp} to avoid permutations. 
The steps for modes $1 < k < d$ are given in \cref{alg:gemm}. 
\begin{algorithm}
	\caption{MTTKRP-GEMM}\label{alg:gemm}
	\begin{algorithmic}[1]
		\Require $\Y, \{\A^{(m)}\}_{\substack{m=1}}^d, R, k$ 
		\Ensure $\Gk$ 
        
		\State $\Z_R^T = \diag(\lvec)(\A_d \odot \cdots \odot \A_{k+1})^T$
		\State $\C = \Call{reshape}{\Y, [I_L \cdot I_k, I_R]} \Z_R$ \Comment{GEMM}
		\State $\Tn{C} = \Call{reshape}{\C, [I_L, I_k, R]}$ \Comment{tensorize the matrix}
        
		\State $\Z_L^T = \diag(\lvec)(\A_{k-1} \odot \cdots \odot \A_1)^T$
        
		\State $\Gk(\ell,j) = \sum_{q=1}^{I_L} \Tn{C}(q, \ell, j) \Z_L(q, j)$ \Comment{parallelize with einsum, GEMM, etc.}
	\end{algorithmic}
\end{algorithm}
\Call{MTTKRP-GEMM}{} has a storage complexity of ${\mathcal{O}}(R(I_L + I_R))$ as the left and right factor matrices can be stored in the same temporary memory space. However, for modes-$1,d$, the storage cost of $\mathcal{O}(R(\prod_{m=1}^{d} I_n / \min \{I_1, I_d\}))$ is the same cost as forming $\Z^T$ explicitly in \cref{eq:krp}.

\subsection{MTTKRP-ELEM}\label{sec:elem}
The simplest attempt to parallelize \cref{eq:elem} is to assign every thread within a team $p_x$ to a tensor element $\Y(i)$ without enforcing any particular league-wise structure. 
This algorithm, \Call{MTTKRP-ELEM}{}, is described in \cref{alg:elem}. Tensor indices $i$ are given by the team offset plus thread rank in line~\ref{line:elem:idx}, hence the tensor reads in line~\ref{line:elem:elem} are packed on CPUs and coalesced on GPUs. 
Reading $\{ \Ak{m} \}_{m=1}^d$ and writing to $\Gk$ requires a conversion of the linear index $i$ to a multi-index $(i_1, \dots, i_d)$ in line~\ref{line:elem:multi} with the function $\Call{ind2sub}{\Y, i}$, which uses costly integer divisions. 
Lines~\ref{line:elem:vecstart}-\ref{line:elem:vecend} are the vector parallelism over columns of the factor matrices $\{ \Ak{m} \}_{m=1}^d$. 
Line~\ref{line:elem:simd} ensures that each vector-thread $p_y$ processes \emph{at least} $F$ columns, and the while loop on lines~\ref{line:elem:loopstart}-\ref{line:elem:loopend} increases the number of columns processed by each thread to $R/(b_y \times F)$. 
Line~\ref{line:elem:coa} ensures that the column strides are packed/coalesced for a fixed $p_x$, but this is negated by line~\ref{line:elem:rand} as the multi-index entry $i_m$ is \emph{not coalesced} for neighboring $p_x$; these \emph{pseudo-random} reads of the factor matrices have a \emph{significant negative effect} on memory bandwidth as $R(d-1)$ factor matrices' columns are read per tensor element $\Y(i)$.
Finally, $\Gk$ is \emph{atomically} updated for each $N$ tensor element $R$ times on line~\ref{line:elem:atomic}.

\begin{algorithm}
	\caption{MTTKRP-ELEM}\label{alg:elem}
	\begin{algorithmic}[1]
		\Require $\Y, \{\Ak{m}\}_{\substack{m=1}}^d, R, k, F$
		\Ensure $\Gk$ 
		\LeagueRange{$l \in [N/b_x]$}
			\TeamRange{$p_x \in b_x$}
				\State $i \gets l \times N/b_x + p_x$ \label{line:elem:idx}
				\State $y_i \gets \Y[i]$ \Comment{packed/coalesced reads} \label{line:elem:elem}
				\State $(i_1, \dots, i_d) \gets \Call{ind2sub}{\Y, i}$ \label{line:elem:multi}
				\VecRange{$p_y \in [b_y]$}\label{line:elem:vecstart}
					\State $jj \gets 0$
					\State $\vec\varphi \gets \mathbf{0}$\label{line:elem:simd}
					\While{$jj < R$} \Comment{read column chunks}\label{line:elem:loopstart}
						\State $ j\gets jj + F \times b_y + p_y$
						\State $\vec\varphi \gets x_i \times \bldl[j:j+F]$
						\For{$m = 1, \dots, d$}
							\If{$m \neq k$}
                                
								\State $\vec\varphi \gets \vec\varphi \times \Ak{m}[i_m, j:j+F]$ \Comment{pseudo-random reads}\label{line:elem:rand}
							\EndIf
						\EndFor
						\State $\Call{AtomicAdd}{\Gk[i_k,j:j+F], \vec\varphi}$ \Comment{$\mathcal{O}(NR)$ atomic updates} \label{line:elem:atomic}
					\State $jj \gets jj + F \times b_y$\label{line:elem:coa}
					\EndWhile\label{line:elem:loopend}
				\EndVecRange\label{line:elem:vecend}
			\EndTeamRange
		\EndLeagueRange
	\end{algorithmic}
\end{algorithm}

\subsection{MTTKRP-SUB}\label{sec:mttkrp-slice}
We would like our element-based parallelization scheme to avoid the atomic updates and pseudo-random reads and writes of the \Call{MTTKRP-ELEM}{} algorithm. 
To avoid atomic operations, \cref{eq:elem} informs us to process each mode-$k$ subtensor (i.e., the $d-1$ subtensor formed by fixing mode $k$, see \cref{fig:slice}) $\Tn{S}^{(k)}_n$ independently.

\begin{figure}[h]
    \centering
    \input{tikz/slice}
    \caption{Subtensors of a 3-way tensor $\Y \in \R^{I_1 \times I_2 \times I_3}$ along each mode.}
    \label{fig:slice}
\end{figure}

We construct an algorithm $\Call{MTTKRP-SUB}{}$ that assigns league and team parallelism to each subtensor $\Tn{S}^{(k)}_n$ of $\Y$. 
The $\Call{MTTKRP-SUB}{}$ algorithm is described by \cref{alg:tile} with $N_T=N_S=N/I_k$ (i.e., the number of elements in a given subtensor $\Tn{S}^{(k)}_n$) and modifying the atomic add on line~\ref{line:tile:atomic} to be a \emph{conflict-free} global update.
The algorithm begins by assigning each team to a subtensor $\Tn{S}^{(k)}_n$ in line~\ref{line:tile:slice}. 
Line~\ref{line:tile:anch} then computes an \emph{anchor} multi-index $a$ for the subtensor, in this case given by $a_m = 0 $ for $m \neq k$ and $a_k = n$. 
Lines~\ref{line:tile:vecstart}-\ref{line:tile:vecend} are the vector parallelism over the factor matrices $\{ \Ak{m}\}_{m=1}^d$ columns and \emph{differ} from the vector parallelism in \cref{alg:elem} as each vector-thread $p_y$ processes $N_T$ tensor elements $y_i$ (lines~\ref{line:tile:elemstart}-\ref{line:tile:elemend}).
Line~\ref{line:tile:multi} computes the tensor multi-index $(i_1, \dots i_d)$ as the sum of the anchor multi-index $(a_1, \dots, a_d)$ and the $\Call{ind2sub}{\Tn{s}^{(k)}_n, ii}$. 
Note that the resulting multi-index $i$ is length $d$ (i.e., the same length of a multi-index for $\Y$), and the $k$th entry of the multi-index $i_k = n$.
For different subtensors $S_n^{(k)}$ and $S_m^{(k)}$, the multi-indices differ only in their $k$th element, hence in practice we implement a special $\Call{ind2sub}{\Tn{s}^{(k)}_n, ii}$ function that uses \emph{the same} precomputed offsets for each subtensor, avoiding the need to explicitly form the subtensor and compute costly inner loop integer divisions used in the classic implementation.
Line~\ref{line:tile:sub2ind} takes the resulting multi-index and calls a function $\Call{sub2ind}{\Y, (i_1, \dots, i_d)}$ built on cheap integer additions and multiplications to get the tensor linear index $i$.
Line~\ref{line:tile:league} assigns all $I_k$ subtensors to a Kokkos team, forming a league grid of size $(I_k, b_y)$.
The inner loop over tensor elements in a subtensor allows the factor matrices columns to be cached and reused (line~\ref{line:tile:reuse}) in the computation of the \emph{element Hadamard product} $\vec\varphi = y_i \prod_{m \neq k} \Ak{m}(i_m, j)$.
The element Hadamard product $\vec\varphi$ is then added to the \emph{subtensor Hadamard product} $\vec\pi$ on line~\ref{line:tile:acc}.
After the inner subtensor loop terminates on line~\ref{line:tile:vecend}, the subtensor Hadamard products $\vec\pi$ are written to $\Gk(i_k,j:j+F)$ \emph{without atomics}.

\subsection{MTTKRP-TILE}
\begin{figure*}[h]
    \centering
    \includegraphics[width=\textwidth]{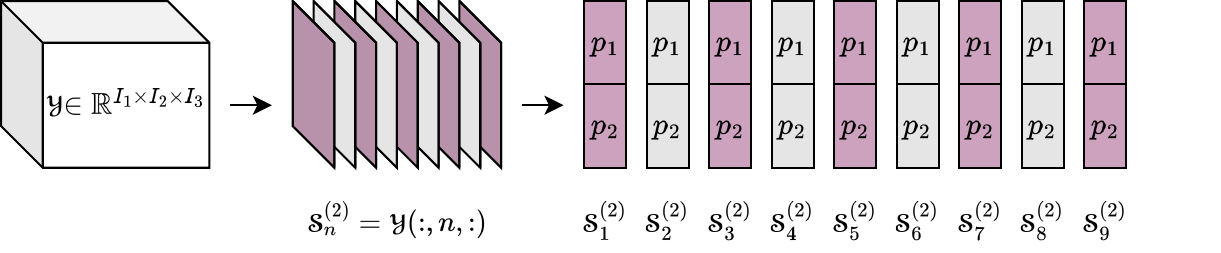}
    \caption{A tensor $\Y \in \R^{I_1 \times I_2 \times I_3}$ decomposed into mode-$2$ subtensors $\Tn{S}_n^{(2)}$. 
    	Each subtensor is then assigned to a Kokkos team of two threads $p_1, p_2$, where each team computes a \emph{tile Hadamard product} and atomically reduces their local contributions to $\Gk(n,j)$. 
	In this instructive example we have one team per subtensor, but in practice we allow for more than one team per subtensor. However, the map between teams and tiles is always bijective.}
    \label{fig:tile}
\end{figure*}

The \Call{MTTKRP-SUB}{} algorithm has the advantage over \Call{MTTKRP-ELEM}{} in that it \emph{avoids atomic conflicts} and has \emph{beneficial cache blocking} for columns of the factor matrices $\{ \Ak{m} \}_{m=1}^d$. 
However, \Call{MTTKRP-SUB}{} has $\mathcal{O}(N_S)$ more serial operations, which can lead to a \emph{significant loss of parallelism} for MTTKRPs on modes with large subtensors (i.e., a mode-$k$ MTTKRP where $I_k$ is large).
We attempt to remedy this loss of parallelism by assigning teams to tiles within a subtensor (see \cref{fig:tile}). 
Given a subtensor $\Tn{S}_k^{(n)}$, a tile partition, defined by a \emph{tile volume} $N_T$, portions the subtensor as $\{\Tn{S}_n^{(k)}((t-1)N_T : tN_T) \}_{t=1}^{N_S/N_T}$, where the set elements $\Tn{S}_n^{(k)}((t-1)N_T : tN_T)$ are called \emph{tiles}.

We call the resulting algorithm that parallelizes over tiles \Call{MTTKRP-TILE}{} and describe it in \cref{alg:tile}. 
The algorithm is controlled by the tile volume hyperparameter $N_T$, which controls \emph{both} the length of the serial accumulation of the tile Hadamard product in lines~\ref{line:tile:elemstart}-\ref{line:tile:elemend} and the number of atomic operations in line~\ref{line:tile:atomic}, which is $\mathcal{O}((N_S/N_T) R)$. 
Parallelism is controlled by the tile volume; line~\ref{line:tile:league} launches $N/N_T$ teams, i.e., a smaller $N_T$ launches more teams.
Note that setting a tile size of $N_T =N_S= N/I_k$ and removing the atomic stipulation on line~\ref{line:tile:atomic} reverts the algorithm to \Call{MTTKRP-SUB}{}, while setting a tile size of $N_T=1$ reverts the algorithm to \Call{MTTKRP-ELEM}{}, albeit with more structure in the team assignments.
We refer to the discussion in \cref{sec:mttkrp-slice} for a detailed discussion of the repeated pseudo-code in \cref{alg:tile}.

\begin{algorithm}
	\caption{MTTKRP-TILE}\label{alg:tile}
	\begin{algorithmic}[1]
		\Require $\Y, \{\Ak{m}\}_{\substack{m=1}}^d, R, k, F$; tile volume $N_T$
		\Ensure $\Gk$ 
		\State $N_S \gets N/I_k$ \Comment{slice volume}
		\LeagueRange{$l \in [N/N_T]$} \label{line:tile:league}\Comment{parallelism over tiles}
			\TeamRange{$p_x \in [N_T]$ } \Comment{thread reuse $N_T>b_x$}
			\State $n \gets \Call{intdiv}{l\times b_x + p_x, N_S / N_T}$\label{line:tile:slice} \Comment{$\Tn{S}^{(k)}_n$}
				\State $t \gets \Call{intmod}{l\times b_x + p_x, N_S / N_T}$
				\State $(a_1, \dots, a_m) \gets \Call{ind2sub}{\Tn{S}_n^{(k)}[(t-1)N_T], 0}$\label{line:tile:anch}
				\State $jj \gets 0$
				\VecRange{$p_y \in [b_y]$}\label{line:tile:vecstart}
					\State $\vec\pi \gets \mathbf{0}$ \Comment{init. tile Hadamard product}
					\State $\vec\varphi \gets \mathbf{0}$ \Comment{init. element Hadamard product}
					\While{$jj < R$}
						\For{$ii \in [N_T]$}\label{line:tile:elemstart} 
							\State $(i_1, \dots, i_d) \gets (a_1, \dots, a_d) + \Call{ind2sub}{\Tn{S}_n^{(k)}[(t-1) N_T], ii}$\label{line:tile:multi}
							\State $i \gets \Call{sub2ind}{\Y, (i_1, \dots, i_d)}$\label{line:tile:sub2ind}
							\State $y_i \gets \Y[i]$ \Comment{re-read tensor element}
							\State $j \gets jj + F \times b_y + p_y$
							\State $\vec\varphi \gets y_i \times \lvec[j:j+F]$
							\For{$m = 1, \dots, d$}
								\If{$m \neq k$}
									\State $\vec\varphi \gets \vec\varphi \times \Ak{m}[i_m, j:j+F]$ \Comment{columns are reused}\label{line:tile:reuse}
								\EndIf
							\EndFor
							\State $\vec\pi \gets \vec\pi + \vec\varphi$ \Comment{update tile product}\label{line:tile:acc}
						\EndFor\label{line:tile:elemend}
						\State $\Call{AtomicAdd}{\Gk[n,j:j+F], \vec\pi}$ \Comment{$\mathcal{O}((N_S/N_T) R)$ atomic updates}\label{line:tile:atomic}
						\State $jj \gets jj +  F \times b_y$
					\EndWhile
				\EndVecRange\label{line:tile:vecend}
			\EndTeamRange
		\EndLeagueRange
	\end{algorithmic}

\end{algorithm}

%% file: tikz/slice.tex
\resizebox{\columnwidth}{!}{%
\begin{tikzpicture}[
    x={(0.90cm, 0.00cm)},
    y={(0.00cm, 0.90cm)},
    z={(-0.38cm,-0.28cm)},
]

\def\W{1.8}   %
\def\H{1.4}   %
\def\D{1.8}   %

\colorlet{sliceI}  {black!10}
\colorlet{sliceII} {black!10}
\colorlet{sliceIII}{black!10}

\newcommand{\boxoutline}{%
}

\begin{scope}[shift={(0cm,0cm)}]
  \foreach \yy in {0.0, 0.35, 0.70, 1.05}{%
    \fill[sliceIII, opacity=0.50]
        (0,\yy,0)--(\W,\yy,0)--(\W,\yy,\D)--(0,\yy,\D)--cycle;
    \draw[sliceIII!65!black]
        (0,\yy,0)--(\W,\yy,0)--(\W,\yy,\D)--(0,\yy,\D)--cycle;
  }
  \boxoutline
  \node[font=\small, align=center] at (0.5*\W, -0.95, 0.5*\D)
      {$\mathcal{S}_n^{(1)} = \mathcal{Y}(n,:,:)$\\[1pt]
       {}};
\end{scope}

\begin{scope}[shift={(3.30cm,0cm)}]
  \foreach \xx in {0.0, 0.45, 0.90, 1.35}{%
    \fill[sliceII, opacity=0.50]
        (\xx,0,0)--(\xx,\H,0)--(\xx,\H,\D)--(\xx,0,\D)--cycle;
    \draw[sliceII!65!black]
        (\xx,0,0)--(\xx,\H,0)--(\xx,\H,\D)--(\xx,0,\D)--cycle;
  }
  \boxoutline
  \node[font=\small, align=center] at (0.5*\W, -0.95, 0.5*\D)
      {$\mathcal{S}_n^{(2)} = \mathcal{Y}(:,n,:)$\\[1pt]
       {}};
\end{scope}

\begin{scope}[shift={(6.60cm,0cm)}]
  \foreach \zz in {0.0, 0.45, 0.90, 1.35}{%
    \fill[sliceI, opacity=0.50]
        (0,0,\zz)--(\W,0,\zz)--(\W,\H,\zz)--(0,\H,\zz)--cycle;
    \draw[sliceI!65!black]
        (0,0,\zz)--(\W,0,\zz)--(\W,\H,\zz)--(0,\H,\zz)--cycle;
  }
  \boxoutline
  \node[font=\small, align=center] at (0.5*\W, -0.95, 0.5*\D)
      {$\mathcal{S}_n^{(3)} = \mathcal{Y}(:,:,n)$\\[1pt]
       {}};
\end{scope}
\end{tikzpicture}
}

%% file: sections/results.tex
\section{Numerical results}
We study the performance and portability of the different MTTKRP algorithms described in \cref{sec:methods}.
Our numerical experiments are performed on two tensors with randomly generated values based on relevant scientific simulations \cite{goal_decomp}: the 2D compressible tearing mode tensor \cite{tear2, tear1} $\Tn{A} \in \R^{401 \times 201 \times 12 \times 501}$, about $3.7$ GB in double precision, 
and the 3D island coalescence tensor \cite{island} $\Tn{B} \in \mathbb{R}^{129 \times 129 \times 129 \times 12 \times 39}$, about $7.5$ GB. 
Our experiments use CP-ranks $R=10, 500, 1000, 2000$. 

\subsection{Experimental setup}
Each algorithmic variant is implemented in the publicly available GenTen library\footnote{\url{https://github.com/sandialabs/GenTen}} built atop Kokkos for performance portability.
Our numerical experiments are performed on cluster nodes of the Hops (OpenMP/CUDA) and El Dorado (HIP) machines at Sandia National Labs.
Each Hops node has two 3.8 GHz Intel Xeon Platinum 8480+ CPUs and four Nvidia H100 SXM5 GPUs; each El Dorado node has four 1.8 GHz AMD 4th Gen EPYC CPUs and four AMD Instinct MI300A GPUs.
For the CPUs, we set $b_x=b_y=1$ (i.e., we only use league parallelism), use GenTen heuristics (see \cite{genten} Tables 3,4) for selecting $F$ based on $R$, and compile with the
Intel MKL LAPACK and BLAS and architecture specific vectorization flags set by Kokkos.
We also set the environment variables $\texttt{OMP\_NUM\_THREADS}=56$, $\texttt{OMP\_PROC\_BIND}=\texttt{close}$, and $\texttt{OMP\_PLACES}=\texttt{threads}$. 
On GPUs, we set $b_x=128/b_y$ and choose $b_y$ and $F$ based on $R$ using the aforementioned GenTen heuristics.

\subsection{Performance analysis}\label{sec:perf}
We use a model $\flops = NRd$ to measure the flop count of the MTTKRP as for each $N$ tensor elements, we do $R(d-1)$ multiplications and $R$ additions. 
Memory bandwidth is measured for each element-based algorithmic variant using a $0$-cache model $\mops_0$, an infinite-cache model $\mops_\infty$, and a $0,\mathbf{LM}$-cache model $\mops_{0,\mathbf{LM}}$, where $\mathbf{LM}$ is the middle level of cache on a device, i.e., L1 cache on a GPU and L2 cache on a CPU.
The $\mops_\infty$ model is designed for compute bound kernels where reading data from cache is cheap compared to the computations, the $\mops_0$ assumes data movement is expensive (i.e., the cost of reading from global memory is always incurred), and $\mops_{0,\mathbf{LM}}$ attempts to split the difference by modeling a cache like $\mops_\infty$ but assigning a cost to accessing said cache.
For the matrix-free mode-$k$ MTTKRPs, the infinite-cache model is given by $\mops_\infty = s_\flops(N + R\sum_{n} I_n)$, where $s_\flops$ is the size in bytes of the floating-point type, $N$ is the size of the tensor, $R\sum_{n \neq k}I_n$ is the size of reading all but the $k$th factor matrix, and $RI_k$ is the size of the output matrix.
The $0$-cache model is given by $\mops_0 = s_\flops(N + NR(d-1) + (N_S/N_T) I_k R)$ as the whole tensor must be read in, \emph{every} tensor element $N$ must read \emph{each} factor matrix column $R$ for $d-1$ factor matrices, and each $R$ columns of the output matrix are updated $N_S/N_T$ times for each $k$ subtensors.
The ${0,\mathbf{LM}}$-cache model assumes that the tensor and output matrices cannot be cached as they \emph{are not} read repeatedly, but the factor matrices \emph{are} cached, yielding the model $\mops_{0, \mathbf{LM}} = s_\flops (N + N_S/N_T(I_kR)) + \frac{1}{l} s_\flops(NR(d-1))$, where $l$ is the ratio between $\mathbf{LM}$ bandwidth and device memory bandwidth.

We use these models to predict the total time $T_0 = \flops / \tau_\flops + \mops_0 \tau_\mops$, $T_\infty =\flops / \tau_\flops + \mops_\infty / \tau_\mops$, and $T_{0,\mathbf{LM}} = \flops / \tau_\flops + \mops_{0, \mathbf{LM}} / \tau_\mops$, where $\tau_\flops$ is the peak machine measured in flops and $\tau_\mops$ is the peak machine bandwidth measured in read/write operations per second. We define arithmetic intensity to be $\ai_i = \flops/\mops_i$, $i \in \{0, \infty, (0,\mathbf{LM})\}$, and we say that a model is compute bound if $\ai_i > \tau_\flops/\tau_\mops$ for a chosen $i$.
Given a wall-time $t$, the throughput of the model is measured as $\thru = \flops/t/1024^3$ and the memory bandwidths for each model are given by $\band_\infty = \mops_\infty/t$, $\band_{0,\mathbf{LM}} = \mops_{0,\mathbf{LM}}/t$, and $\band_0 = \mops_0/t$.

Peak throughput $\tau_\flops$,  bandwidth $\tau_\mops$, and $\mathbf{LM}$ bandwidth ratio $l$ for each device is reported in \cref{tab:peak}.
For the H100, throughput and bandwidth are taken from NVIDIA's whitepaper\footnote{\url{https://www.NVIDIA.com/en-us/data-center/h100/}} and we calculate the peak L1 cache bandwidth as the number of SMs $\times$ the L1 transfer bytes/cycle $\times$ the clock rate, which is $123 \times 128 \times 1.98 / 10^3 \approx 33$TB/s, where the L1 transfer bytes per clock cycle is taken from NVIDIA's hopper tuning guide\footnote{\url{https://docs.NVIDIA.com/CUDA/hopper-tuning-guide/index.html}}. Hence, for the H100, we set $l=10$.
For the MI300A, throughput, bandwidth, L1 transfer bytes per clock cycle, and max clock rate are taken from AMD's whitepaper\footnote{\url{https://www.amd.com/content/dam/amd/en/documents/instinct-tech-docs/data-sheets/amd-instinct-mi300a-data-sheet.pdf}} and we use the same formula to calculate peak L1 cache bandwidth as $228\times128\times2.1/10^3\approx61$TB/s; for the MI300A, we set $l=12$.
Throughput for the 8480+ is taken from Intel's APP metrics sheet\footnote{\url{https://www.intel.com/content/www/us/en/support/articles/000005755/processors.html}} 
while memory bandwidth and L2 cache bandwidth are measured with the STREAM triad benchmark \cite{stream}, where we use a 2GB array to measure memory bandwidth and a 0.6MB array to measure the L2 cache bandwidth.

\begin{table}[h]
\centering
\caption{Peak compute throughput $\tau_\flops$, memory bandwidth $\tau_\mops$, and cache bandwidth ratio $l$ for each device. }
\begin{tabular}{ lccc  }
 \toprule
 Device & $10^{12}\tau_\flops$ & $1024^{4}\tau_\mops$ & $l$ \\
 \midrule
 8480+  & 2.28 & 0.12 & 8\\
 H100   & 34 & 3.35  & 10\\
 MI300A & 61.3 & 5.3   & 12 \\
 \bottomrule
\end{tabular}
\label{tab:peak}
\end{table}

We analyze the efficacy of our performance models in \cref{tab:model}. Our results show that the cache effects play a significant role for each algorithmic variant: the $T_\infty$ prediction is much too optimistic for each variant.
The \Call{MTTKRP-ELEM}{} algorithm fails to surpass even the 0-cache predicted time $T_0$ for either device. As discussed in \cref{sec:elem}, this can be attributed to the uncoalesced repeated reads of the factor-matrices. 
The \Call{MTTKRP-SUB}{} algorithm also fails to surpass $T_0$ on the GPU, though it does so on the CPU. We hypothesize that this discrepancy is due to the much larger cache size of the CPU compared to the GPU.
For \Call{MTTKRP-SUB}{} on the CPU and \Call{MTTKRP-TILE}{} on both devices, the $0$-cache model is far too pessimistic while the $\infty$-cache model is too optimistic, indicating that the algorithms achieve (to some extent) the desired caching effects on the repeated reads of the factor matrices. 
Our empirical results for these methods align most with the $0, \mathbf{LM}$-cache model, achieving at least $90\%$ of the predicted time on the CPU, $40\%$ of the predicted time on the H100 GPU, and $10\%$ of the predicted time on the MI300A GPU.
We surmise that the higher accuracy of our models on the CPU is due to the fact that we use empirically measured peak bandwidth, while on the GPU we use the vendor reported ideal world numbers.

\begin{table*}[h]
\centering
\caption{Execution time $T$ in seconds measured against analytical performance model times $T_0, T_{0, \mathbf{LM}}, T_\infty$ for the tensors $\Tn{A}, \Tn{B}$ on the OpenMP, CUDA and HIP execution spaces. The tile size $N_T$ for the \emph{MTTKRP-TILE} algorithm is selected via the heuristic in \cref{eq:heuristic}. The CP-rank is fixed to $R=32$. }
\begin{tabular}{ clgcccgcccgccc}
 \toprule
 && \multicolumn{4}{c}{OpenMP}&\multicolumn{4}{c}{CUDA}&\multicolumn{4}{c}{HIP}\\
 \cmidrule{3-14}
 && $T$ & $T_0$ & $T_{0, \mathbf{LM}}$ & $T_\infty$ & $T$ & $T_0$ & $T_{0,\mathbf{LM}}$ & $T_\infty$ & $T$ & $T_0$ & $T_{0, \mathbf{LM}}$ & $T_\infty$\\
 \midrule
	\multirow{3}*{$\Tn{A}$}&ELEM & 10.394 & 3.817 & 1.349 & 0.056 & 1.823  & 0.138 & 0.047 & 0.003 & 2.668 & 0.087 & 0.028 & 0.002\\
			       &SUB & 0.543  & 2.877 & 0.409 & 0.056 & 8.528  & 0.104 & 0.013 & 0.003 & 16.267 & 0.066 & 0.007 & 0.002\\
			       &TILE & 0.526  & 2.955 & 0.487 & 0.056 & 0.043  & 0.110 & 0.019 & 0.003 & 0.063 & 0.069 & 0.011 & 0.002\\
 \midrule
	\multirow{3}*{$\Tn{B}$}&ELEM & 25.167 & 9.876 & 3.054 & 0.130 & 3.924  & 0.356 & 0.105 & 0.007 & 7.995 & 0.225 & 0.063 & 0.004\\
			       &SUB & 1.137  & 7.927 & 1.105 & 0.130 & 23.032 & 0.286 & 0.035 & 0.007 & 47.401 & 0.181 & 0.019 & 0.004\\
			       &TILE & 1.517  & 8.414 & 1.592 & 0.130 & 0.120  & 0.304 & 0.052 & 0.007 & 0.202 & 0.188 & 0.026 & 0.004\\
 \bottomrule
\end{tabular}
\label{tab:model}
\end{table*}

\subsection{Experiments}
\subsubsection{Selecting the tile size}\label{sec:tilesize}
Our first goal is to select the heuristic tile volume parameter $N_T$ in the \Call{MTTKRP-TILE}{} algorithm. 
\cref{tab:width} presents the numerical performance of a sweep over \emph{tile-widths} $\sqrt[d-1]{N_T} \in \{2,4,6,8,10,12\}$ for the tearing-mode and island coalescence tensors for a fixed $R=32$.
On the GPU, we find that for the 4-way tensor $\Tn{A}$, a tile size of $216$ is optimal, while for the 5-way tensor $\Tn{B}$, a tile size of $256$ is optimal.

On an H100, our maximum occupancy with our standard block configuration of $(4,32)$ is $64$ warps per SM. 
Nsight Compute reports our achieved occupancy to be $60\%$, but even if we assume a more conservative occupancy of $50\%$, we have $32$ tiles per SM, with each tile rereading $8\times256$ bytes of 
factor matrices entries, or $64$KB of information per SM, $25\%$ of the total L1 cache, a good target given that the L1 cache handles read, write, and atomic instructions on a GPU.  

On the 8480+, we find that a tile width of $12$, i.e., the size of the minimum dimension, is optimal. 
Our parallel CPU implementation assigns one tile per core, and given that each core has 2MB of L2 cache, this means that the factor matrices take up at most $\sim 0.5\%$ of the L2 cache.
However, our implementation assumes that the tile size is \emph{regular}, i.e., that $\sqrt[d-1]{N_T} \in \mathbb{Z}$. As such, performance decreases when $\sqrt[d-1]{N_T} > \min_{n=1, \dots, d}I_n$.

To impose tile regularity and cache awareness, we use the simple heuristic 
\begin{equation}\label{eq:heuristic}
	N_T = \min \left\{\left(\sqrt[d-1]{\frac{s_\mathbf{LM}/4}{s_\flops(c/2) }} \right)^{d-1}, \min_{n=1,\dots,d}\{I_n\}^{d-1} \right\},
\end{equation}
where $s_\mathbf{LM}$ is the size in bytes of the middle level of cache (i.e., L1 cache (per SM) on a GPU, L2 cache (per core) on a CPU), and $c$ is the maximum number of tiles per cache unit (i.e., SM/core). 
We find that this heuristic is valid for most $R$.

\begin{table*}[h]
\centering
\caption{Sweep over tile widths $\sqrt[d-1]{N_T}$  for the \emph{MTTKRP-TILE} algorithmic variant with $R=32$ for each tensor $\Tn A, \Tn B$ on both execution spaces, with performance in $\thru$ (higher is better) averaged over all modes. These results motivate a heuristic to choose $N_T$ given by \cref{eq:heuristic}.}
\begin{tabular}{ rcccggg  }
 \toprule
 &\multicolumn{6}{c}{\Call{MTTKRP-TILE}{} $\thru$} \\
 \midrule
	\multirow{2}{*}{{$\sqrt[d-1]{N_T}$}}&\multicolumn{3}{c}{Tearing mode tensor $\Tn{A}$}&\multicolumn{3}{g}{Island coalescence tensor $\Tn{B}$} \\
 \cmidrule{2-7}
& OpenMP & CUDA & HIP & OpenMP & CUDA & HIP\\
 \midrule
 2   & 32.9           & 253.3           & 168.0& 51.4           & 585.1          & 296.6\\
 4   & 89.9           & 1313.6          & 801.8& 99.0           & \textbf{1232.2}& \textbf{836.3}\\
 6   & 98.9           & \textbf{1328.3} & \textbf{910.1}& 106.8          & 1173.4         & 743.1\\
 8   & 98.1           & 1103.7          & 789.2& 104.9          & 944.1          & 580.8\\
 10  & 96.5           & 947.3           & 674.2& 103.3          & 884.4          & 477.7\\
 12  & \textbf{99.6}  & 1230.0          & 808.3& \textbf{110.8} & 948.8          & 475.9\\
\bottomrule
\end{tabular}
\label{tab:width}
\end{table*}

\subsection{Memory impact of MTTKRP variants}\label{sec:mem}

\begin{figure}[h]
    \centering
    \includegraphics[width=\columnwidth]{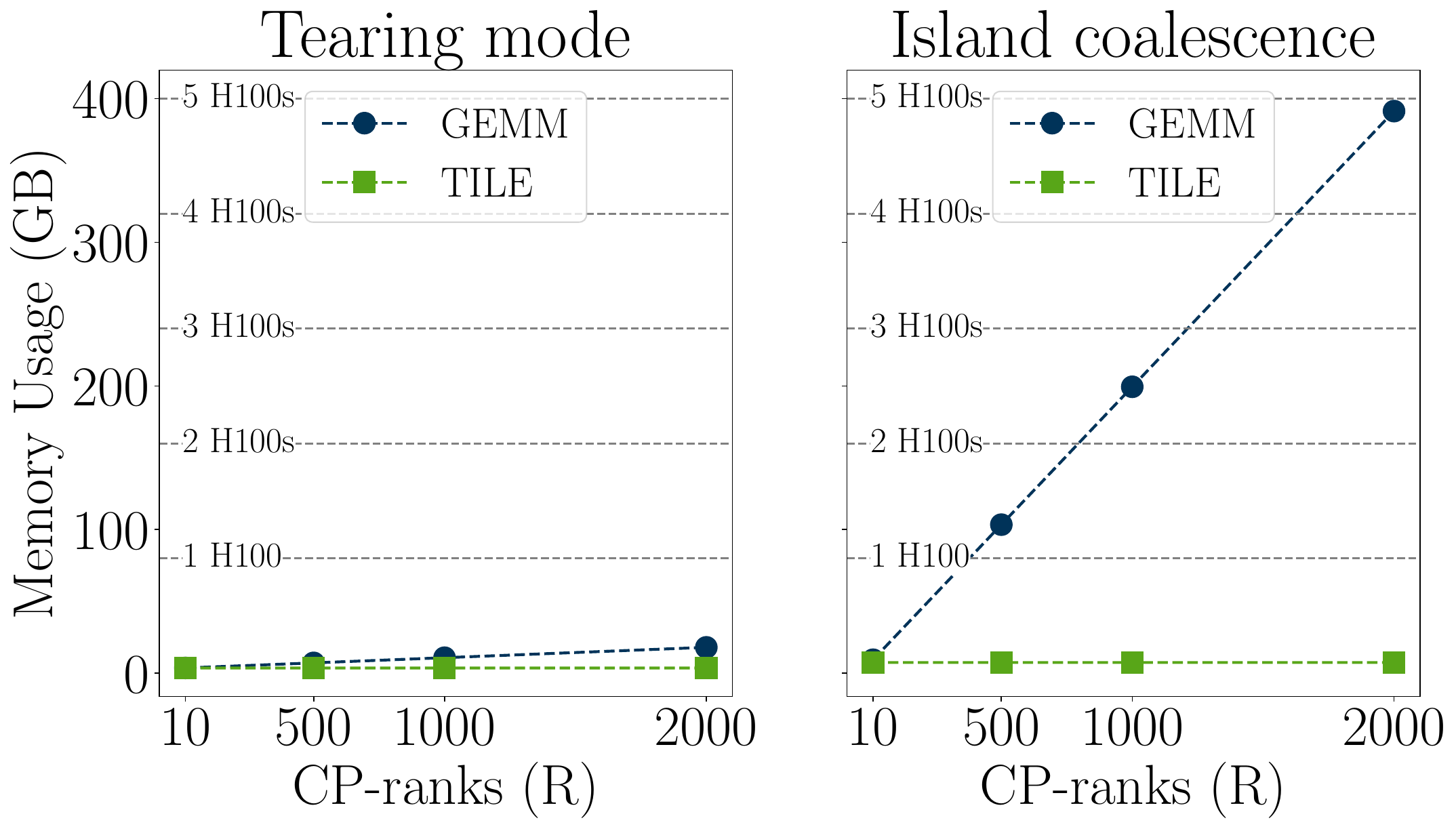}
    \caption{Worst case memory usage for the \emph{GEMM} and \emph{TILE} MTTKRP variants for the tearing mode tensor $\Tn{A}$ and the island coalescence tensor $\Tn{B}$ for different CP-ranks. The plot is annotated with lines showing the minimum number of NVIDIA H100 GPUs required to incur the storage costs. }
    \label{fig:mem}
\end{figure}

As discussed in \cref{sec:methods-gemm}, forming the left and right Khatri-Rao matrices $\Z_L^T$ and $\Z_R^T$ can incur a large storage cost. 
This cost dominates the total memory required for the mode-$k$ \Call{MTTKRP-GEMM}{} algorithm, which is given by $\mops_\infty^\mathrm{GEMM} = N + R(I_L + I_R + I_k)$. 
It is important to note that the storage cost of the \Call{MTTKRP-GEMM}{} algorithm relies heavily on the \emph{initial shape} of a given tensor.
For example, the maximum memory usage over all modes $k$ for the island coalescence tensor $\Tn{B}$ for $R=2000$ is 391GB, but if the initial tensor shape was permuted to be $129 \times 12 \times 129 \times 39 \times 129$ (best case), the memory usage would be 123GB, while if the tensor was permuted to be $129\times129\times129\times39\times12$ (worst case), the memory usage would be 1255GB. 
However, especially in the context of in situ decomposition of scientific simulation data \cite{goal_decomp, cpu_mttkrp}, reshaping tensors can eliminate multidimensional relationships between tensor entries while tensor transpositions can be expensive and impractical. 
\cref{fig:mem} shows the maximum memory usage over all modes $k$ for the \Call{MTTKRP-GEMM}{} and \Call{MTTKRP-TILE}{} algorithms for the tearing mode tensor $\Tn{A}$ and the island coalescence tensor $\Tn{B}$.
The $y$ axis of the plot denotes the \emph{minimum} number of 80GB H100 GPUs needed for storage; however, the distribution protocol in \cref{sec:dist} does not evenly assign memory to MPI ranks, hence \emph{in practice} more GPUs may be required for a given problem size.   

\begin{figure}[h]
    \centering
    \input{images/roofline_h100.pgf}
    \caption{Roofline plot for the tearing mode tensor $\Tn{A}$ with rank $R=2000$ on the NVIDIA H100. The $\mathbf{m}_\infty$ model is used for the \emph{GEMM} MTTKRP variant and the $\mathbf{m}_{0, \mathbf{LM}}$ model for the \emph{ELEM} and \emph{TILE} MTTKRP variants.}
    \label{fig:roofline}
\end{figure}

\subsubsection{Single-node performance}\label{sec:single}
\begin{figure*}[h]
    \centering
    \includegraphics[width=\textwidth]{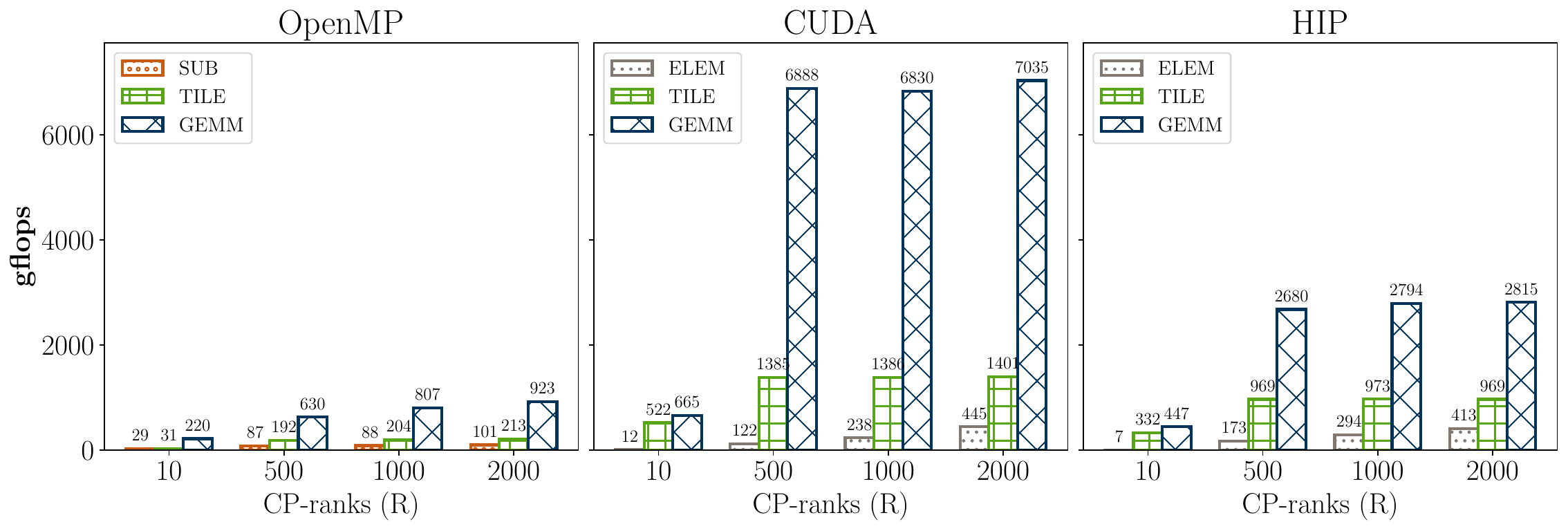}
    \caption{Tearing mode tensor $\Tn{A}$ average (over all modes) MTTKRP performance (higher is better) for different algorithmic variants on different execution spaces.}
    \label{fig:single}
\end{figure*}

The linear-in-rank memory scaling of the \Call{MTTKRP-GEMM}{} algorithm described in \cref{sec:methods-gemm} motivates the use of the tearing-mode tensor $\Tn{A}$ to study single-node performance of the different MTTKRP variants. 
\cref{fig:roofline} shows a roofline analysis the ELEM, TILE, and GEMM kernels on $\Tn{A}$ with $R=2000$ on the H100 GPU; the GEMM kernel is sufficiently compute bound while the matrix-free kernels are memory bound and close to the roofline.
\cref{fig:single} reports the average MTTKRP performance, measured in $\thru$, over all modes, for the OpenMP, CUDA, and HIP execution spaces. 
We compare the SUB, TILE, and GEMM variants on the OpenMP space and the ELEM, TILE, and GEMM variants on the CUDA and HIP spaces, where the tile size $N_T$ is chosen by \cref{eq:heuristic}. 
We omit the ELEM variant for OpenMP and the SUB variant for CUDA/HIP given their poor performance shown in \cref{tab:model}. 
For $R>10$, \Call{MTTKRP-TILE}{} achieves at least $20\%$ of \Call{MTTKRP-GEMM}{}'s performance, a favorable result given \Call{MTTKRP-TILE}{}'s memory footprint.
Additionally, for $R > 10$, \Call{MTTKRP-TILE}{} performs at least $2\times$ faster than the SUB variant on the OpenMP space, at least $3\times$ faster than ELEM on the CUDA space, with a peak speedup of $11\times$ for $R=500$, and at least $2\times$ faster that ELEM on the HIP space, with a peak speedup of $6\times$ for $R=500$.
These speedups reinforce the necessity for data reuse and reduced atomic contention for element-based parallelization approaches.

\subsubsection{Comparison against distributed MTTKRP-GEMM}\label{sec:dist}
The disparities of memory costs between the \Call{MTTKRP-TILE}{} and \Call{MTTKRP-GEMM}{} algorithms for the island coalescence tensor $\Tn{B}$ (see \cref{fig:mem}) motivate us to compare running the \Call{MTTKRP-TILE}{} algorithm on a single GPU against running \Call{MTTKRP-GEMM}{} algorithm on (the required) multiple GPUs.
Distributed MTTKRPs use the communication pattern introduced in \cite{dist_mttkrp} whose communication overhead consists of an initial tensor redistribution and all-reduce communication to combine contributions across processors.

\begin{figure}[h]
    \centering
    \includegraphics[width=\columnwidth]{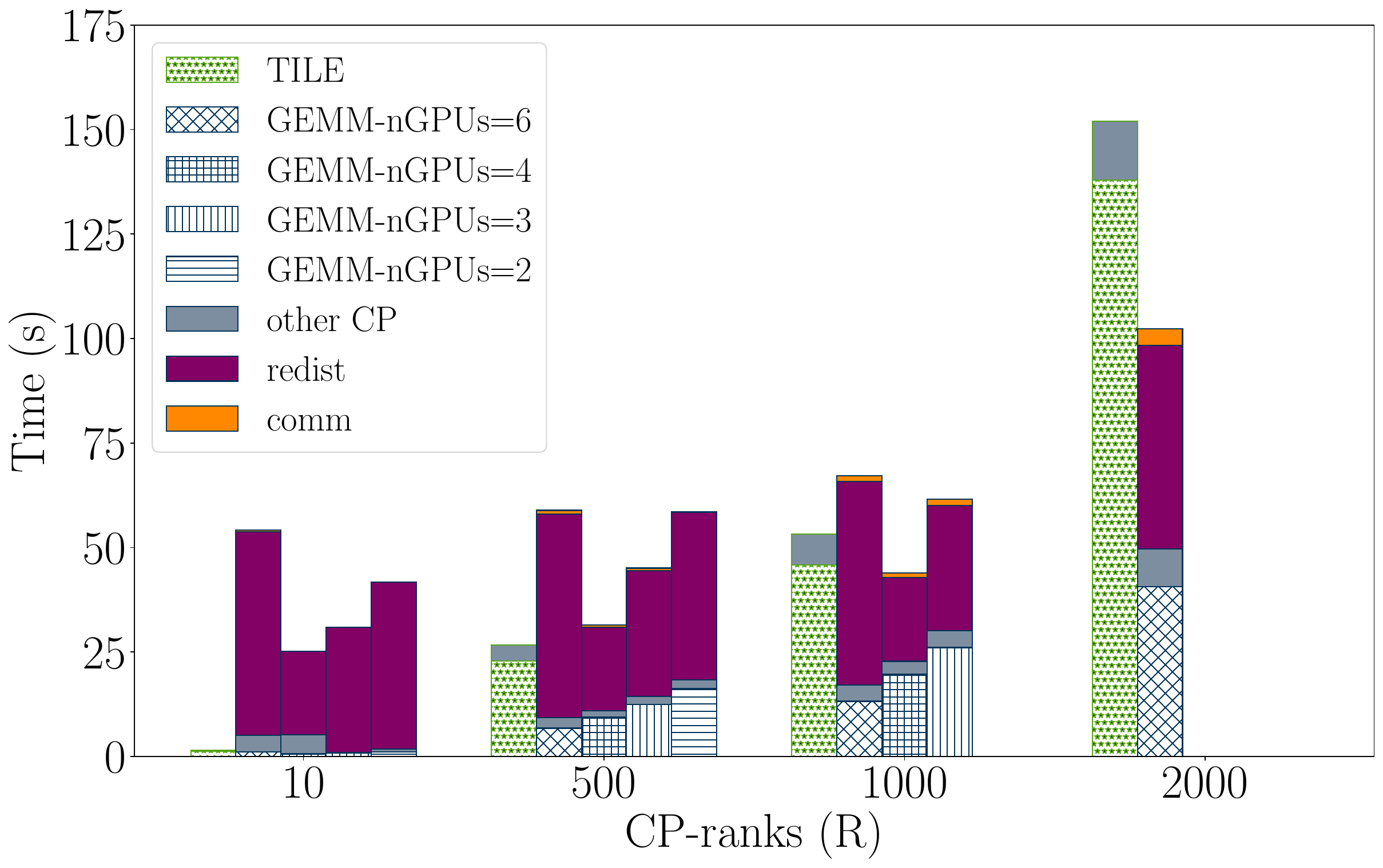}
    \caption{Time (lower is better) for CP-ALS on the island coalescence tensor $\Tn{B}$ with \emph{MTTKRP-TILE} algorithm on a single H100 compared with \emph{MTTKRP-GEMM} on one H100 per process. MTTKRP time is denoted with hatches, other CP-ALS operations (i.e., inner-product, Cholesky solve, etc.) by a grey fill, tensor redistribution time with a purple fill, and MTTKRP communication time with an orange fill.}
    \label{fig:dist}
\end{figure}

\cref{fig:dist} compares the execution time of the CP-ALS algorithm on $\Tn{B}$ with tolerance $10^{-4}$ over ranks $R=10,500,1000,2000$ on a single device when using the \Call{MTTKRP-TILE}{} algorithm and on multiple devices when using the \Call{MTTKRP-GEMM}{} algorithm.
While tensor and factor matrices fit snugly into memory for the TILE variant, \Call{MTTKRP-GEMM}{} requires at least six H100s for rank 2000, three H100s for rank 1000, and two H100s for rank 500 (findings in line with our lower bound memory model). 
Note that the Hops machine requires two nodes to distribute across six GPUs, with the multi-node communication bandwidth drastically increasing tensor redistribution and MTTKRP communication time.  
When $R=2000$, MTTKRP-TILE-based CP-ALS on a single GPU is able to achieve 67\% of the performance of MTTKRP-GEMM-based CP-ALS run on the minimum six H100s. 
For $R=1000$, four H100s are required to beat the MTTKRP-TILE-based CP-ALS and \emph{at least} five are required to beat the MTTKRP-TILE-based CP-ALS for $R=500$.
For all tested ranks, tensor redistribution is the most expensive overhead incurred in the distributed CP-ALS, dwarfing the all-reduce communication time.  

%% file: sections/conclusion.tex
\section{Conclusion}
We introduce fast and memory-efficient algorithms for matrix-free dense mode-$k$ MTTKRPs together with detailed performance analysis and evaluations on CPUs and GPUs. 
We extend the open-source GenTen package with our methods and demonstrate that state-of-the-art matrix-based MTTKRPs need many GPUs to compute the same tensor decomposition that, when replaced with our MTTKRP, can be computed on a single device.
As a general rule, \emph{we recommend that a practitioner use the GEMM algorithm when product matrices fit into memory and highly optimized BLAS-3 rutines are available; otherwise, use the TILE algorithm with our heuristic for selecting $N_T$}.

Our methods have limitations.
\cref{fig:single} shows that we have yet to achieve the efficiency of the matrix-based MTTKRP-GEMM algorithm when the Khatri-Rao product matrix fits in device memory.
\cref{tab:model} motivates us to further improve our caching strategies to achieve a higher percentage of the predicted infinite-cache time $T_\infty$.
Our roofline analysis (\cref{fig:roofline}) shows that our TILE algorithm needs to move into the compute-bound regime to achieve higher throughput. Strategies to increase the arithmetic intensity  include GEMM-like tilings of the output matrix: a 2D block tiling over the output matrix with shared memory caching, thread and warp tiling, and league-size/team-size auto-tuning---in short, mimicking GEMM without forming the product matrices.
Additional optimizations include a Hilbert curve or ALTO-like \cite{alto} ordering of the input tensor to increase cache locality beyond the standard row-major layout.
The optimizations stated above are future work.